# Pseudo-Measurement Enhancement in Power Distribution Systems

Tao Xu, Kaiqi Wang, Jiadong Zhang
Key Laboratory of Smart Grid of
Ministry of Education
Tianjin University
Tianjin, China
taoxu2011@tju.edu.cn

Ji Qiao, Zixuan Zhao
China Electric Power Research Institute
Beijing, China

Hong Zhu, Kai Sun
State Grid Nanjing Electric Power
Company
Nanjing , China

*Abstract*—With the rapid development of smart distribution networks (DNs), the integrity and accuracy of grid measurement data are crucial to the safety and stability of the entire system. However, the quality of the user power consumption data cannot be guaranteed during the collection and transmission process. To this end, this paper proposes a low-rank tensor completion model based on CANDECOMP/PARAFAC decomposition (CPD-LRTC) to enhance the quality of the measurement data of the DNs. Firstly, the causes and the associate characteristics of the missing data are analyzed, and a third-order standard tensor is constructed as a mathematical model of the measurement data of the DN. Then, a completion model is established based on the characteristics of measurement data and the low rank of the completion tensor, and the alternating direction method of multipliers (ADMM) is used to solve it iteratively. Finally, the proposed model is verified through two case studies, the completion accuracy, the computational efficiency, and the memory usage are compared to traditional methods.

## I. Introduction

With the rapid development of the power industry, the demand for power quality and supply stability has been continuously increasing. The distribution network (DN) is no longer limited to meet the basic load demand, but has entered a more refined intermediate development stage. Its core goal has shifted to ensure that users have a highly reliable power supply [1]. The widespread integration of distributed energy resources (DERs) and electric vehicles have brought profound changes to the operating model of the DN. As the link directly connected to users, the operating status of the DN is critical to the reliability of the power supply. Therefore, real-time monitoring and accurate estimation of the DN's status have become important. However, the low quality of measurement data reduces the system's observability and reliability. Data loss during the collection and transmission process can result from delays, communication interference, sensor failures, and other factors [2]. Thus, the use of pseudo-measurement data enhancement methods is crucial in supplementing missing data and improving the observability and reliability of the DN's data [3].

Research on missing data handling across multiple fields have gained wide spread attention, such as traffic flow analysis, image processing, and signal analysis, various data imputation methods have been proposed. Traditional approaches, such as linear prediction, mean and expectation-maximization imputation [4], perform well under low missing data rates but experience significant performance drops and increased model complexity as missing rates rise. Recently, deep learning has demonstrated significant potential in missing data imputation and data augmentation. A self-attention-based imputation for time series (SAITSs) method to recover missing data in distributed phasor measurement units was proposed in [5], while [6] introduced a data-augmentation-based convolutional denoising autoencoder (DACDA) to restore missing building energy consumption data. Although these methods achieve strong performance, they rely on large amounts of high-quality historical data, making them less effective in data-scarce scenarios. Moreover, their ability to handle periodic characteristics and high-peak fluctuations remains limited, posing challenges in highly volatile power systems and energy consumption data. To overcome the limitations of existing methods, low-rank tensor completion (LRTC) technology has emerged, as its multidimensional structure can efficiently handle complex missing data [7], effectively leveraging low-rank constraints for data imputation without requiring historical data for training. While existing LRTC methods achieve high completion accuracy at high missing data rates, they require extensive iterative computations and face high storage and computational costs, which limits their large-scale application.

To address the issue of missing measurement data in DNs, this paper proposes a CPD-LRTC-based method for data completion, considering the inherent low-rank nature of the power measurement data, the electrical correlations between measurement parameters, and the spatial correlations among electricity users. First, a three-order tensor model of user electricity data in the DN is established. Then, tensor CP decomposition is performed, and the resulting factor matrices are used to replace the traditional tensor unfolding matrices, significantly reducing computational complexity and memory usage while ensuring completion accuracy and efficiency. The completed missing data serves as pseudo-measurement data for the DN, enhancing its observability.

The rest of this article is organized as follows. Section II analyzes the causes of DNs' missing measurement and the associate data characteristics. Section III introduces tensor theory, formulates the proposed model, and utilizes the ADMM method to solve it. Section IV presents case studies and comparative analyses of the algorithm. The final section provides conclusions and the future research directions.

## II. Analysis of Distribution Network Measurement

### A. Analysis of Causes of Missing Measurement Data

Smart meters have been widely deployed in DNs, but data collection and transmission still face missing data issues in actual operation. The main causes include the following.

*1) **Communication failures**:* Communication faults between smart meters and the information collection system can disrupt data transmission, resulting in the inability to collect user data timely and accurately, which in turn affects the system's monitoring and control functions.

*2) Remote terminal failures:* Given the real-time requirements of the meter collection system, failures of remote terminals may lead to data loss, which cannot be recovered through backtracking.

*3) Power failure events:* Large-scale power outages, load shedding, and other events can cause data loss or anomalies, affecting data quality.

*4) System design flaws:* Inadequate system design may result in missing or incomplete data.

*5) Environmental Factors:* Uncontrollable factors such as magnetic interference or extream weather conditions can cause abnormal data from smart meters, increasing the difficulty of ensuring data quality.

The completeness and reliability of electric measurement data is crucial for the stable operation and efficient management of the power system. Therefore, effective data imputation is required to improve data quality.

### B. Analysis of Measurement Data Characteristics

DN measurement data refers to the data collected from users, including parameters such as current, voltage, power factor, and active power. The main characteristics of this data are as follows:

*1) Randomness:* Due to each user's unique daily routine and electricity usage habits, active power data exhibits significant randomness, as shown in Fig.1.

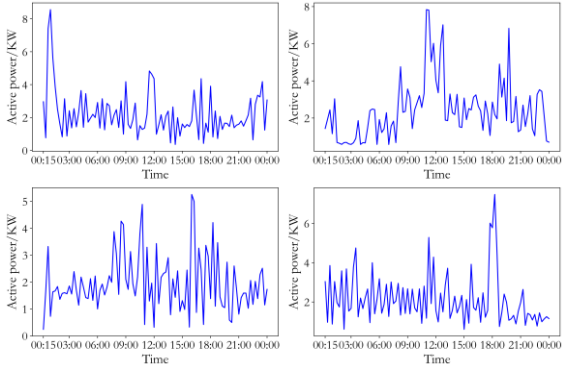

Fig. 1. The active power profiles of different users on the same day

*2) Periodicity:* The electricity usage patterns of residents are relatively stable, resulting in a clear periodic variation in the power consumption profile. As shown in Fig.2, the electricity consumption behavior of a single user typically exhibits periodic patterns.

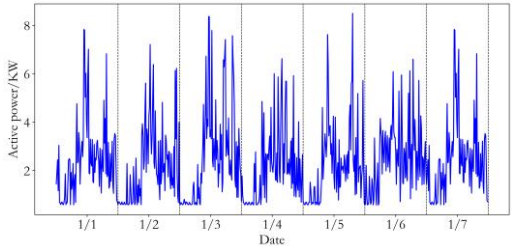

Fig. 2. The active power profile of one user over seven consecutive days

*3) Spatial Correlation:* In a DN, multiple electricity users are closely connected through the network within the feeder area, resulting in significant spatial correlation between power loads.

*4) Electrical Correlation:* The measurement parameters in a DN exhibit electrical correlation: voltage $U$, current $I$, active power $P$, and power factor $\cos\varphi$ all of which follow the power conservation principle. As shown in Fig.3, when a reactive load is activated, both $P$ and $I$ increase simultaneously, while $\cos\varphi$ decreases, demonstrating the synchronous response characteristics in the dynamic changes of the circuit. The relationship between $U$, $I$, $P$ and $\cos\varphi$ is given by (1).

$$P = UI \cos\varphi \quad (1)$$

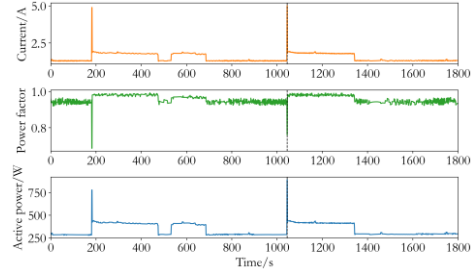

Fig. 3. Current, power factor, and active power data profiles of a smart meter for a specific time period

## III. MODEL OF CPD-LRTC

### A. Related Tensor Computation Operations

*1) Tensor Matrixization:* Matrixization, also known as unfolding, is the process of rearranging the elements of an N-dimensional array in a specific order to construct a matrix structure. In this paper, tensors are denoted by uppercase italic letters, such as $\mathcal{X}$. An $N^{th}$ order tensor can be expressed as $\mathcal{X} \in \mathbb{R}^{I_1 \times I_2 \cdots \times I_N}$, where $I_n$ represents its $n^{th}$ dimension. The matrix obtained by unfolding tensor X along its $n^{th}$ dimension is referred to as $X_{(n)}$. The unfolding operation is expressed as:

$$f(\mathcal{X}, n) = X_{(n)} \in \mathbb{R}^{I_n \times (I_1 \times \cdots \times I_{n-1} \times I_{n+1} \times \cdots \times I_N)} \quad (2)$$

The mapping relationship from tensor element $\mathcal{X}(i_1, i_2, ..., i_N)$ to matrix element $X_{(n)}(i_n, j)$ is expressed as:

$$j = 1 + \sum_{k=1, k \neq n}^{N} (i_k - 1) \prod_{m=1, m \neq n}^{k-1} I_m \quad (3)$$

*2) CP Decomposition of Tensor with Missing Values:* For a tensor $\mathcal{X} \in \mathbb{R}^{I_1 \times I_2 \times \cdots \times I_N}$ with missing values, a weighted least squares model with missing values is proposed as:

$$\min_{U_n} \|\mathcal{W} * (\mathcal{X} - U_1 \circ U_2 \circ \cdots \circ U_N)\|_F^2 \quad (4)$$

where, $*$ represents the Hadamard product, and $U_n \in \mathbb{R}^{I_n \times R}$ is called tensor factor matrices, which comes from the combination of rank-one tensors, such as $U_1 = [a_1, a_2, ..., a_R]$, as shown in Fig.4. $R > 0$ is the rank of the CP decomposition. $\mathcal{W}$ is a non-negative weight tensor of the same size as $\mathcal{X}$, which stores the positional information of the observable and missing measurements in the tensor.

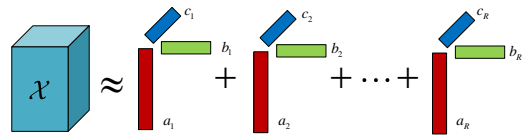

Fig. 4. Illustration of the CP decomposition model of a third-order tensor with rank $R$

## B. The Construction of Measurement Data Tensor

In the low-voltage DN, smart meters collect electricity consumption data from each user including user information, usage date, time period, and parameters. This study constructs a third-order tensor $\mathcal{T} \in \mathbb{R}^{I_1 \times I_2 \times I_3}$, where $I_1$ represents the number of consecutive days of data collection, $I_2$ denotes sampling points within 24 hours, $I_3$ indicates the number of users or measurement parameters. The tensor structure is shown in Fig.5. The focus of this study is to detect the positions of missing data in the tensor $\mathcal{T}$ and use the proposed model to complete the missing data.

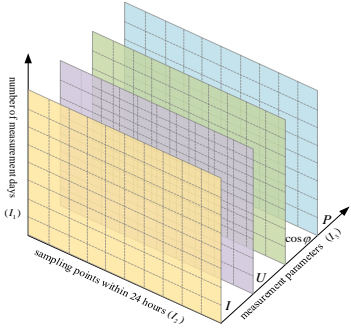

Fig. 5. Measurement data third-order tensor structure diagram

## C. CPD-LRTC Method

Based on the previously analyzed characteristics of measurement data, it is clear that the user electricity data captured by smart meters exhibits strong spatial and temporal correlations, which further suggests that the constructed three-order tensor structure tends to be low-rank or close to low-rank. As an extension of matrices in higher-dimensional spaces, the concept of low-rank is also introduced in tensors. The rank of a tensor $\mathcal{X}$ is defined as $rank(\mathcal{X})$. Therefore, the objective function of the model is:

$$\begin{cases} \min_{\mathcal{X}} rank(\mathcal{X}) \\ s.t.\ \mathcal{X}_\Omega = \mathcal{T}_\Omega \end{cases} \quad (5)$$

where, $\mathcal{X}$ and $\mathcal{T}$ are the tensors to be completed and the original value tensors, respectively. $\Omega$ represents the set of observable data positions in the tensor.

Given that computing the $rank(\mathcal{X})$ is an NP-hard problem, the $rank(\mathcal{X})$ can be replaced with a weighted sum of the ranks of its mode-n unfolding. The unfolded matrices of $\mathcal{X}$ are derived by unfolding the tensor along the day mode, the sampling point mode, and the user or measurement parameter mode. These correspond to the periodicity, temporal correlation, and spatial or electrical correlation features of the DN measurement data, respectively. This approach enables the integration of features from different modes, thereby achieving high-precision completion of missing data. However, due to the large memory consumption of the mode-unfolded matrices and the lengthy iterative computation time, further optimization is necessary.

Let $X_{(n)}$ be the mode-n unfolding matrix of the tensor $\mathcal{X}$ with rank $r$, and let $U_n$ be the corresponding CP decomposition factor matrix, where $n = 1, \ldots, N$. It has been proven by a theorem that the rank of $U_n$ is an upper bound of the rank of $X_{(n)}$, that is,

$$rank(X_{(n)}) \leq rank(U_n), \quad \forall n = 1, \ldots, N. \quad (6)$$

Therefore, $U_n$ can be used to replace $X_{(n)}$, then,

$$\begin{cases} \min_{\mathcal{X}} \sum_{i=1}^{3} \alpha_i rank(U_n) \\ s.t.\ \mathcal{X}_\Omega = \mathcal{T}_\Omega, \quad \mathcal{X} = U_1 \circ U_2 \circ U_3 \end{cases} \quad (7)$$

where, $U_n \in \mathbb{R}^{I_n \times R}$, $n = 1, 2, 3$, and $R$ is the upper bound of the $rank(\mathcal{X}) = r$, which is an integer, i.e., $R \geq r$.

Due to the discrete nature of the rank, (7) is a non-convex optimization problem. It can be transformed into a convex optimization problem by replacing the rank function with the trace norm as a convex surrogate, as shown below:

$$\begin{cases} \min_{\mathcal{X}} \sum_{i=1}^{3} \alpha_i \|U_n\|_* \\ s.t.\ \mathcal{X}_\Omega = \mathcal{T}_\Omega, \quad \mathcal{X} = U_1 \circ U_2 \circ U_3 \end{cases} \quad (8)$$

The augmented Lagrange function is established to remove the equality constraints in (8), and its relaxed expression is:

$$\begin{cases} \min_{\mathcal{X}} \sum_{i=1}^{3} \alpha_i \|U_n\|_* + \frac{\lambda}{2} \|\mathcal{X} - U_1 \circ U_2 \circ U_3\|_F^2 \\ s.t.\ \mathcal{X}_\Omega = \mathcal{T}_\Omega. \end{cases} \quad (9)$$

(9) simultaneously performs tensor CP decomposition and LRTC, requiring the execution of singular value decomposition (SVD) only on smaller factor matrices. Therefore, the proposed model significantly reduces computational complexity and the iteration time.

## D. Model Solution

The objective function in (9) is difficult to solve due to the presence of interdependent matrix nuclear norm terms. When optimizing the sum of multiple matrix nuclear norms, these matrices share common terms and cannot be optimized independently. By splitting these interdependent terms, they can be solved independently. Introducing auxiliary matrices $M_n \in \mathbb{R}^{I_n \times R}, n = 1, 2, 3$, the equivalent formulation of (9) is:

$$\begin{cases} \min_{\mathcal{X}} \sum_{i=1}^{3} \alpha_i \|M_n\|_* + \frac{\lambda}{2} \|\mathcal{X} - U_1 \circ U_2 \circ U_3\|_F^2 \\ s.t.\ \mathcal{X}_\Omega = \mathcal{T}_\Omega,\ M_n = U_n,\ n = 1, 2, 3 \end{cases} \quad (10)$$

(10) is a constrained optimization problem. To solve it, its augmented Lagrange function can be developed as follows:

$$\mathcal{L}_\mu(U_n, M_n, \mathcal{X}, Y_n) = \sum_{i=1}^{3} \alpha_i \|M_i\|_* + \frac{\lambda}{2} \|\mathcal{X} - U_1 \circ U_2 \circ U_3\|_F^2 \\ + \sum_{n=1}^{3} (\langle Y_n, M_n - U_n \rangle + \frac{\mu}{2} \|M_n - U_n\|_F^2) \quad (11)$$

where, $Y_n \in \mathbb{R}^{I_n \times R}$ is the Lagrange multiplier matrix, where $n = 1, 2, 3$; $\langle \cdot \rangle$ denotes the matrix inner product; $\|\cdot\|_F$ is the matrix norm; $\lambda > 0$ is the regularization parameter; and $\mu > 0$ is the penalty parameter.

Based on the ADMM framework, an iterative scheme is proposed: continuously minimize $\mathcal{L}_\mu$ in $(U_n, M_n, \mathcal{X})$, and then update $Y_n$. This approach simplifies the complex optimization problem in (11) into solving four subproblems iteratively.

When solving the $U_n$ subproblem, by fixing $M_n$, $Y_n$, and $\mathcal{X}$, the original problem is transformed into:

$$U_n^{k+1} = \min_{U_n} \frac{\lambda}{2} \| \mathcal{X} - U_1 \circ U_2 \circ U_3 \|_F^2 + \sum_{n=1}^{3} \frac{\mu^k}{2} \left\| U_n - M_n^k - \frac{Y_n^k}{\mu^k} \right\|_F^2 \quad (12)$$

The solution for $U_n$ is:

$$U_n^{k+1} = (\lambda X_{(n)}^k B_n^T + \mu^k M_n^k + Y_n^k)(\lambda B_n B_n^T + \mu^k I)^{-1} \quad (13)$$

where, $B_n = (U_N^k \odot \cdots U_{n+1}^k \odot U_{n-1}^{k+1} \odot \cdots U_1^{k+1})^T$.

When solving the $M_n$ subproblem, fixing $U_n$, $Y_n$, and $\mathcal{X}$, the original problem is transformed into the following form:

$$M_n^{k+1} = \min_{M_n} \| M_n \|_* + \frac{\mu^k}{2} \left\| M_n - U_n^{k+1} - \frac{Y_n^k}{\mu^k} \right\|_F^2 \quad (14)$$

The closed-form solution of (14) is obtained by solving it using the singular value thresholding (SVT) algorithm.

$$M_n^{k+1} = SVT_{\alpha_n/\mu^k}(U_n^{k+1} - Y_n^k / \mu^k) \quad (15)$$

where, $SVT_{\alpha_n/\mu^k}(\cdot)$ refers to performing a soft thresholding operation with a threshold of $\alpha_n / \mu^k$.

Similarly, when solving the $\mathcal{X}$ subproblem, by fixing $U_n$, $M_n$ and $Y_n$, and applying the Karush-Kuhn-Tucker (KKT) conditions, the solution for the tensor $\mathcal{X}$ is obtained as:

$$\mathcal{X}^{k+1} = \mathcal{P}_\Omega(\mathcal{T}) + \mathcal{P}_{\Omega^c}(U_1^{k+1} \circ \cdots \circ U_N^{k+1}) \quad (16)$$

where, $\Omega^c$ is the complement of $\Omega$, which represents the index set of the unobserved entries in the tensor, and $\mathcal{P}_\Omega(\cdot)$ is the projection operator.

The iterative form of the Lagrange multiplier matrix $Y_n$ is:

$$Y_n^{k+1} = Y_n^k + \mu^k(M_n^{k+1} - U_n^{k+1}), \quad n = 1, 2, 3 \quad (17)$$

The alternating optimization process for the four local subproblems is performed as follows. When the convergence condition $\| \mathcal{X}^{k+1} - \mathcal{X}^k \|_F / \| \mathcal{X}^0 \|_F \leq \varepsilon = 10^{-4}$ is met, the output $\mathcal{X}$ can be the completed.

## IV. CASE STUDIES

### A. Datasets

The case study data used in this paper is from two datasets. Dataset 1 contains active power data from 114 apartments in a low-voltage DN in the UK, used as a multi-user single measurement dataset. Dataset 2 is from an apartment in Delhi, India, containing 73 consecutive days of smart meter data, used as a single-user multi-measurement dataset.

For Case 1, a month of active power data without missing values is selected to construct a three-order tensor. For Case 2, 21 consecutive days without missing values are selected, using $P$, $U$, $I$, and $\cos\varphi$ to form a three-order tensor, and random missing values are introduced for testing.

### B. Analysis of Missing Data Imputation Results for DN Measurements

#### 1) Multi-User Single Measurement Data

Fig.6 and Fig.7 show the completion results of active power values for Apartment 1 with a missing rate of 30% on a single day and a week. The results indicate that the algorithm can effectively complete the missing data, with the completed blue profile fitting well with the original red curve, and the filled data aligning with the actual power consumption.

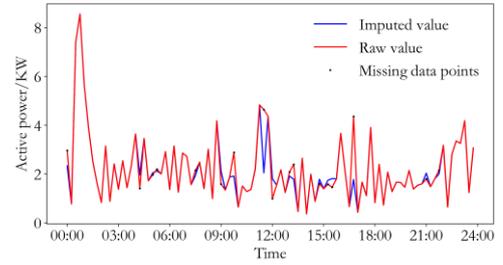

Fig. 6. Comparison of active power profiles before and after imputation for Apartment 1 on a specific day with a 30% missing rate.

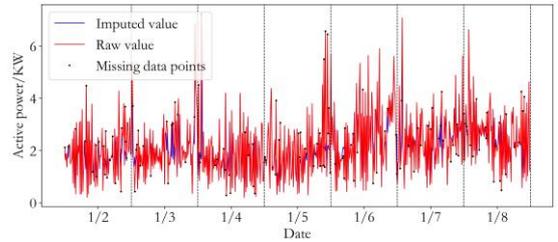

Fig. 7. Comparison of active power profiles before and after imputation for Apartment 1 on a specific week with a 30% missing rate.

To simulate random missing data, the complete dataset is processed with missing rates from 10% to 90%. Table I presents the relative squared error (RSE) and computation time for missing data completion by the CPD-LRTC model compared to the traditional LRTC algorithm at various missing rates.

TABLE I. CASE 1 RSE AND COMPUTATION TIME

| Method | HaLRTC | | CPD-LRTC | |
|---|---|---|---|---|
| Missing rate/% | RSE/% | Time/s | RSE/% | Time/s |
| 10 | 13.22 | 7.011 | 13.01 | 4.089 |
| 20 | 19.08 | 6.861 | 18.89 | 3.877 |
| 30 | 23.69 | 6.904 | 22.33 | 3.728 |
| 40 | 27.45 | 6.751 | 25.99 | 3.878 |
| 50 | 31.20 | 6.792 | 29.75 | 4.141 |
| 60 | 34.56 | 6.908 | 31.95 | 3.918 |
| 70 | 38.05 | 7.011 | 34.02 | 4.667 |
| 80 | 41.96 | 6.926 | 38.25 | 5.636 |
| 90 | 47.03 | 6.686 | 42.69 | 4.815 |

When the missing rate is below 50%, the RSE of the completed data remains within 30%. Thus, it can be inferred that this algorithm effectively completes multi-user, single-measurement electricity data when the missing rate does not exceed 50%. Given that the missing rate in practical applications generally stays below this threshold, the proposed method offers significant advantages in data completion, enabling efficient and accurate data recovery.

#### 2) Single-User Multiple Measurement Data

For Case 2, prior to the formal completion, pre-filling can be performed using the electrical correlation between multiple measurements, as described in (1), to add valid information to the tensor. The missing data is then completed using the proposed algorithm. The completion error and computation time are shown in Table II.

TABLE II. CASE 2 RSE AND COMPUTATION TIME

| Method | HaLRTC | | CPD-LRTC | |
|---|---|---|---|---|
| Missing rate/% | RSE/% | Time/s | RSE/% | Time/s |
| 10 | 0.73 | 72.56 | 0.75 | 19.25 |
| 20 | 2.32 | 66.28 | 2.53 | 22.46 |
| 30 | 5.95 | 68.55 | 5.87 | 26.34 |
| 40 | 11.63 | 90.61 | 9.52 | 30.52 |
| 50 | 19.42 | 82.81 | 16.74 | 32.87 |
| 60 | 29.20 | 79.07 | 25.68 | 35.69 |
| 70 | 42.18 | 95.00 | 35.59 | 38.87 |
| 80 | 57.74 | 97.18 | 47.67 | 44.48 |
| 90 | 72.57 | 114.62 | 65.38 | 50.59 |

Table II shows that when the random missing rate of electricity data is less than 60%, the RSE of the completed data is less than 30%, indicating that the proposed algorithm can effectively complete multi-measurement data from smart meters when the missing rate is below 60%. Furthermore, the proposed algorithm completes data faster than the traditional LRTC algorithm at any missing rate, demonstrating higher efficiency and superiority.

To visually demonstrate the effect, Fig.8 show the comparison of the current, voltage, and active power profiles for a particular day at a 30% missing rate. The completed data (in blue) closely fits the original data (in red) and accurately fills in the missing points marked in white. This result fully demonstrates the high efficiency and reliability of the proposed method under data random missing scenarios, and its ability to effectively detect and complete missing data.

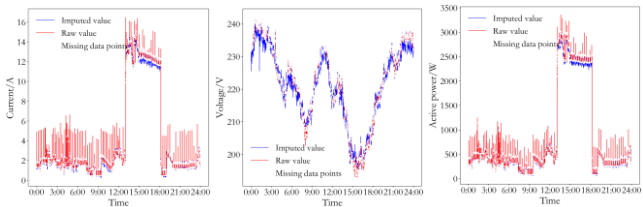

Fig. 8. Comparison of current, voltage and active power profiles before and after imputation for an apartment with a 30% missing rate.

### C. Algorithm Comparison

A comparison of the proposed algorithm with four missing data imputation methods (Kalman filter, cubic spline interpolation, self-attention-based imputation for time series and generative adversarial network) is conducted. Fig.9 shows the trend of the RSE under random missing rates from 10% to 90% for Case 1 and Case 2, respectively, when different imputation algorithms are applied. Overall, the imputation errors for all methods increase as the missing rate rises. However, the CPD-LRTC and the traditional LRTC consistently outperform the other four methods in terms of lower imputation errors, demonstrating their higher imputation accuracy. Compared to traditional LRTC, the CPD-LRTC has similar imputation errors at low missing rates but significantly lower imputation errors at high missing rates. Tables I and II show that the imputation time of CPD-LRTC is lower than that of LRTC at any missing rate, proving its higher solving efficiency.

Specifically, when the missing rate is below 50%, the CPD-LRTC algorithm performs better in Case 2, with RSE not exceeding 20%. When the missing rate exceeds 50%, the error in Case 2 increases exponentially, demonstrating the effectiveness of electrical correlation at low missing rates. Electrical correlation can effectively pre-fill missing data, improving imputation accuracy. However, at high missing rates, the excessive missing data causes the pre-filling strategy to fail, leading to increased errors. Overall, the correlation significantly enhances imputation accuracy at low missing rates.

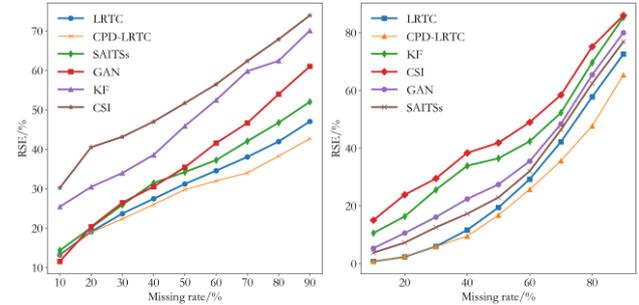

Fig. 9. RSE after imputation for Case 1 and Case 2.

## V. CONCLUSIONS

The method proposed in this paper, based on CPD-LRTC, is suitable for the missing data imputation problem in DN measurement data, and its imputation accuracy outperforms traditional imputation algorithms. Currently, this study only considers the case of random missing data and does not account for non-random missing data. In the future, additional prior constraints can be introduced to handle both random and non-random missing data imputation simultaneously.

ACKNOWLEDGMENT

This work is supported by the State Grid project "Research and application of intelligent computing technology for distribution network status under sparse measurement conditions" (Grant number: 5108-202218280A-2-403-XG).

REFERENCES

[1] H. Huang, H. V. Poor, K. R. Davis, et. al., "Toward resilient modern power systems: From single-domain to cross-domain resilience enhancement," Proceedings of the IEEE, vol. 112, no. 4, pp. 365–398, 2024.

[2] R. M. Jungers, A. Kundu and W. P. M. H. Heemels, "Observability and controllability analysis of linear systems subject to data losses," IEEE Transactions on Automatic Control, vol. 63, no. 10, pp. 3361-3376, 2018.

[3] T. Xu, J. D. Zhang, H. Meng, et. al., "SeqESR-GAN-Based sparse data augmentation for distribution networks," IEEE Transactions on Industrial Informatics, vol. 20, no. 11, pp. 12913-12923, 2024.

[4] X. Y. Miao, Y. Y. Wu, L. Chen, et. al., "An experimental survey of missing data imputation algorithms," IEEE Transactions on Knowledge and Data Engineering, vol. 35, no. 7, pp. 6630-6650, 2023.

[5] S. Nayak, D. Dwivedi, K. Victor, et. al., "Data imputation using self attention based model for enhancing distribution grid monitoring and protection systems," IEEE Transactions on Instrumentation and Measurement, vol. 73, pp. 1-11, 2024.

[6] A. Liguori, R. Markovic, M. Ferrando, et. al., "Augmenting energy time-series for data-efficient imputation of missing values," Applied Energy, vol. 334, pp. 1-18, 2023.

[7] Z. Long, Y. P. Liu, L. X. Chen, et. al., "Low rank tensor completion for multiway visual data," Signal Processing, vol. 155, pp. 153-161, 2018.